# From Explanations to Architecture: Explainability-Driven CNN Refinement for Brain Tumor Classification in MRI

Rajan Das Gupta
American International University-Bangladesh
Department of Computer Science
Dhaka, Bangladesh
18-36304-1@student.aiub.edu

Md Imrul Hasan Showmick
Brac University
Department of Computer Science
Dhaka, Bangladesh
imrul.hasan.showmick@gmail.com

Lei Wei*
Faculty of Psychology
Shinawatra University
Bangkok, Thailand
weileihh21@163.com

Mushfiqur Rahman Abir
American International University–Bangladesh
Department of Computer Science
Dhaka, Bangladesh
20-42738-1@student.aiub.edu

Shanjida Akter
North South University
Department of Computer Science
Dhaka, Bangladesh
shanjida.akter01@northsouth.edu

Md. Yeasin Rahat
American International University–Bangladesh
Department of Computer Science
Dhaka, Bangladesh
20-43097-1@student.aiub.edu

Md. Jakir Hossen
Multimedia University
Department of Computer Science
Malaysia
jakir.hossen@mmu.edu.my

## Abstract

Recent brain tumor classification methods often report high accuracy but rely on deep, over-parameterized architectures with limited interpretability, making it difficult to determine whether predictions are driven by tumor-relevant evidence or spurious cues such as background artifacts or normal tissue. We propose an explainable convolutional neural network (CNN) framework that enhances model transparency without sacrificing classification accuracy. This approach supports more trustworthy AI in healthcare and contributes to SDG 3: Good Health and Well-being by enabling more dependable MRI-based brain tumor diagnosis and earlier detection. Rather than using explainable AI solely for post hoc visualization, we employ Grad-CAM to quantify layer-wise relevance and guide the removal of low-contribution layers, reducing unnecessary depth and parameters while encouraging attention to discriminative tumor regions. We further validate the model's decision rationale using complementary explainability methods, combining Grad-CAM for spatial localization with SHAP and LIME for attribution-based verification. Experiments on multi-class brain MRI datasets show that the proposed model achieves 98.21% accuracy on the primary dataset and 95.74% accuracy on an unseen dataset, indicating strong cross-dataset generalization. Overall, the proposed approach balances simplicity, transparency, and accuracy, supporting more trustworthy and clinically applicable brain tumor classification for improved health outcomes and non-invasive disease detection.

*Corresponding author

## CCS Concepts

• **Do Not Use This Code → Generate the Correct Terms for Your Paper**; *Generate the Correct Terms for Your Paper*; Generate the Correct Terms for Your Paper; Generate the Correct Terms for Your Paper.

**ACM Reference Format:**
Rajan Das Gupta, Md Imrul Hasan Showmick, Lei Wei, Mushfiqur Rahman Abir, Shanjida Akter, Md. Yeasin Rahat, and Md. Jakir Hossen. 2018. From Explanations to Architecture: Explainability-Driven CNN Refinement for Brain Tumor Classification in MRI. In *Proceedings of Make sure to enter the correct conference title from your rights confirmation email (Conference acronym 'XX).* ACM, New York, NY, USA, 8 pages. https://doi.org/XXXXXXX.XXXXXXX

## 1 Introduction

Tumors arise when cells proliferate abnormally and accumulate into a mass, bypassing the tightly regulated cycle of growth, division, and apoptosis observed in healthy tissue. Brain tumors are among the most life-threatening conditions worldwide, driven by a combination of environmental and biological factors, including exposure-related risks (e.g., air pollution) [29] and hereditary susceptibility in a minority of cases, as well as radiation exposure in occupational settings [5]. Accurate diagnosis therefore depends on a clear understanding of brain structure and function [19]. Clinically, tumors are commonly described as benign or malignant: benign lesions are typically slow-growing and less likely to recur after resection, whereas malignant tumors tend to infiltrate or progress rapidly and can cause severe neurological dysfunction without timely intervention [34, 39]. In addition, the World Health Organization grading scheme broadly associates lower grades (I–II) with less aggressive behavior and higher grades (III–IV) with more aggressive disease, with gliomas often representing higher-grade pathology and meningioma/pituitary tumors frequently appearing at lower grades; however, variation in location, texture, and morphology makes robust classification challenging [17, 28].

Brain tumors are frequently examined using medical imaging, with MRI widely regarded as the preferred technique for brain assessment due to its non-invasive nature and its capacity to produce high-contrast, three-dimensional anatomical images, which supports automated analysis [18, 24, 37, 42]. In practice, radiologists aim to localize lesions and distinguish tumor tissue from normal structures before determining tumor type [39]. Manual interpretation becomes difficult at scale because tumor appearance can overlap across classes and varies substantially in size and shape, motivating computer-aided systems to improve efficiency and consistency [40, 41].

To address this, both traditional machine learning and deep learning approaches have been investigated. While classical approaches (e.g., SVM variants) can be effective, modern deep learning has become dominant in MRI-based detection and classification by learning hierarchical representations directly from images, often achieving superior performance [31, 39]. However, deep models are frequently criticized as *black boxes*: their complex architectures hinder clinical interpretability, and without reliable explanations they may over-rely on irrelevant cues (e.g., normal soft tissue) rather than tumor-discriminative regions [9, 22, 25, 33]. Moreover, increasing architectural depth to improve accuracy raises computational cost, complicating real-time clinical deployment. Explainable AI (XAI) addresses this gap by providing post hoc interpretability of model decisions, improving transparency, accountability, and trust in clinical settings.

Despite recent progress, many XAI-enabled brain tumor classifiers still do not *use* explanations to improve the underlying model: explanations are often reported after training, while the model itself remains over-parameterized and potentially prone to attending to spurious regions. Motivated by this limitation, our objective is to leverage XAI not only for interpretation but also for *model refinement*, enabling a less complex architecture that remains accurate and generalizable.

Although explainable artificial intelligence (XAI) is now widely used in medical imaging to make model decisions more transparent, most prior work still treats explanations mainly as *post hoc* visual aids generated after training. Consequently, such explanations rarely influence the design or optimization of the underlying architecture. In contrast, this work treats explainability as an active design signal rather than a passive interpretability layer. Specifically, we leverage Grad-CAM-derived layer-wise relevance maps to guide architectural refinement by identifying and removing convolutional layers that contribute minimally to the final prediction. This explanation-driven refinement enables model simplification while preserving discriminative capability, resulting in a more efficient and interpretable CNN that maintains strong classification performance.

**Contributions.** The main contributions of this study are summarized as follows:

- We employ explainable AI to identify tumor-relevant abnormal regions in brain MRIs, reducing reliance on spurious cues and improving classification reliability.
- We provide an interpretable framework that exposes the reasoning behind predictions via complementary explainers, supporting clinical trust and auditability.
- We translate explanation evidence into model refinement by removing low-contribution layers, yielding a less complex CNN while preserving discriminative performance.
- Our method delivers strong results across both datasets, achieving 98% accuracy on Dataset-1 and 94–95% on the unseen Dataset-2, highlighting its strong generalization capability.

## 2 Related Work

Recent MRI-based brain tumor classification research is dominated by deep learning (DL), with most studies emphasizing accuracy via transfer learning, attention mechanisms, and extensive augmentation. We summarize representative works spanning both *non-explainable* DL pipelines and *XAI-enabled* systems, including a small number of earlier baselines that remain frequently referenced.

### 2.1 Deep Learning Methods Without Explainability

A common trend is to increase representational capacity through attention and transformer-style designs. Alzahrani proposed ConvAttenMixer, combining convolutional mixing with external and self-attention to capture global–local dependencies, reporting strong multi-class accuracy on public MRI benchmarks [3]. Vision Transformer (ViT) variants and ensembles similarly achieve high performance but typically require higher computational cost for training and inference [38]. Alongside attention-based models, many studies rely on CNN backbones and transfer learning. Islam et al. compared transfer learning architectures and reported very high accuracy under augmentation and fine-tuning [16], while Aziz et al. improved DenseNet-based performance using regularization and hyperparameter tuning, highlighting both the benefits and dataset dependence of fine-tuned backbones [7].

Several studies have explored tailored CNN architectures and hybrid learning pipelines. For example, Gupta et al. introduced a lightweight CNN for binary classification and evaluated its performance against widely used pre-trained models. [14]. AlTahhan et al. explored fine-tuned CNNs and hybrid AlexNet–SVM/KNN variants, demonstrating that classical classifiers can complement CNN feature extractors in limited-data regimes [2]. Rasheed et al. combined MRI enhancement (e.g., contrast normalization) with a modified CNN and benchmarked against standard backbones [34]. Others report that direct architectural tailoring can outperform off-the-shelf transfer learning: Özkaraca et al. evaluated CNN/VGG/DenseNet settings and proposed a modified CNN when transfer learning underperformed [30]; Gómez-Guzmán et al. compared a generic CNN against multiple pre-trained models under augmentation [13]. On the smaller Figshare-style benchmark (3264 images), Peng and Liao designed a deeper CNN for four-class classification [32]. Addressing dataset imbalance is another recurring theme: Imam and Alam studied loss functions and oversampling/augmentation strategies to stabilize training under class skew [15]. Beyond DL, classical machine learning remains relevant as a baseline: Arora and Gupta proposed a weighted least-squares twin SVM variant for multi-class classification [4].

Earlier deep CNN baselines are also often cited for establishing feasibility across datasets, despite limited interpretability [6]. Overall, while these approaches deliver strong accuracy, they generally provide limited insight into *why* predictions are made and may still exploit spurious cues.Overall, while these approaches deliver strong accuracy, they generally provide limited insight into why predictions are made and may still exploit spurious cues. Additionally, CNNs are susceptible to noise interference, such as Poisson noise common in imaging data, which can degrade classification performance. Prior work has shown that appropriate activation-function selection, such as ReLU in ResNet50 architectures, can improve robustness under Poisson noise and achieve up to 97% accuracy in multi-class image-classification settings [12].

## 2.2 Explainable AI-Driven Deep Learning Methods

To improve transparency, recent studies integrate explainable AI (XAI) to localize or attribute model evidence. Grad-CAM-based visualization is commonly used to highlight discriminative MRI regions and support qualitative inspection of CNN decisions [20, 27]. SHAP and LIME are also adopted to provide attribution-based explanations at the instance level and to support human verification of predicted classes [8, 10, 25, 33]. Ahmed et al. further applied explanation feedback in an iterative manner, retraining when explanations were deemed inadequate, reflecting an emerging direction toward explanation-aware model development [1]. However, in most XAI-enabled works, explanations are primarily used *post hoc* for visualization rather than as a direct signal to simplify or refine model architecture.

**Gap and motivation.** Prior studies either (i) prioritize accuracy through complex architectures without strong interpretability, or (ii) attach post hoc explainers without using explanation evidence to improve the model itself. This motivates our approach, which leverages XAI not only to justify predictions but also to guide architectural refinement (layer reduction) while preserving accuracy and cross-dataset generalization.

# 3 Materials and Methods

## 3.1 Datasets

We evaluate the multi-class brain tumor classification task on two publicly available brain MRI datasets. The first, the Msoud dataset (Nickparvar, 2021), contains 7,023 grayscale MRI slices stored in JPG format. The second, the NeuroMRI repository, includes 3,264 grayscale MRI slices, also in JPG format. Both datasets comprise four categories: *meningioma*, *glioma*, *pituitary*, and *no-tumor*. Table 1 summarizes the class distributions..

**Table 1: Brain MRI dataset composition by tumor category.**

| Class | Dataset 1 | Dataset 2 |
|---|---|---|
| Meningioma | 1645 | 926 |
| No-tumor | 2000 | 500 |
| Glioma | 1621 | 937 |
| Pituitary | 1757 | 901 |

## 3.2 Preprocessing

MRI slices exhibit variations in resolution, acquisition settings, and field-of-view, which can introduce domain shift and reduce generalization. To standardize inputs and mitigate dataset bias [26], we apply:

1. **Cropping (brain region extraction).** Each image is first converted to grayscale and thresholded to isolate the foreground from the background. Erosion and dilation are then applied to eliminate small artifacts. The largest contour is subsequently detected, and the original image is cropped using the extreme top, bottom, left, and right points of that contour, with a small margin retained around the region of interest.
2. **Normalization.** Cropped images are scaled to the intensity range $[0, 255]$ for consistent contrast.
3. **Resizing.** All images are resized to $224 \times 224 \times 3$ to reduce computation and enforce a fixed input size.

## 3.3 Baseline CNN

Our baseline CNN follows a standard hierarchical feature extractor. The network stacks repeated *Conv(3×3) + ReLU + MaxPool* blocks with increasing filters (8, 16, 32, 64, 128, 256) and `same` padding. A batch normalization layer is applied before average pooling. The resulting feature maps are flattened and passed through two fully connected layers (512 units each, ReLU). A final softmax layer outputs class probabilities for four classes.

## 3.4 Grad-CAM for spatial and layer-wise relevance

To improve the interpretability of model predictions, we employ Gradient-weighted Class Activation Mapping (Grad-CAM) [35], which highlights class-discriminative regions by leveraging gradients propagated to a selected convolutional layer. For reference, conventional CAM uses global average pooling (GAP) to compress

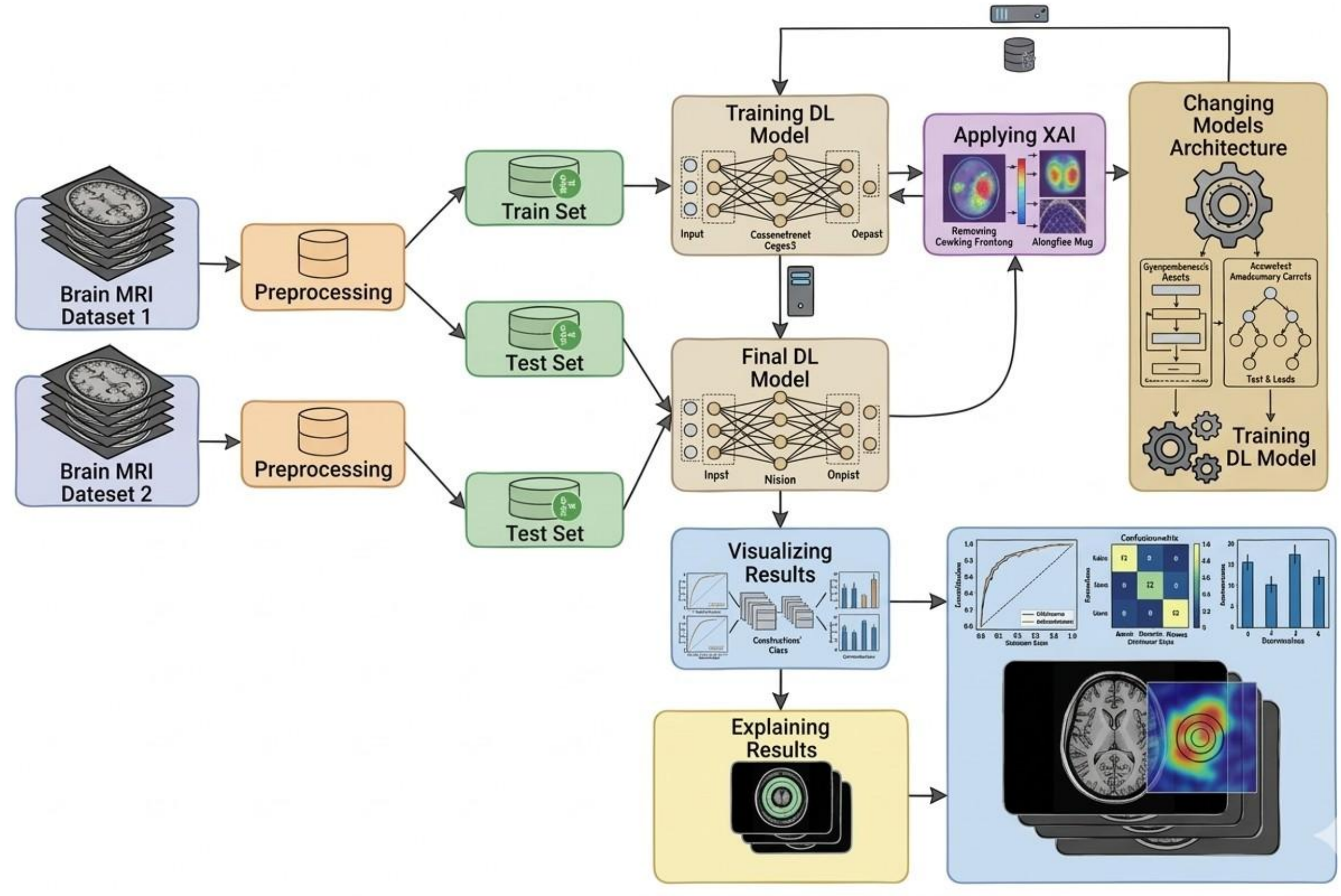

**Figure 1: An XAI-driven framework for brain tumor classification**

a feature map $A^{(\,)} \in \mathbb{R}^{\times}$ into a scalar:

$$g = \frac{1}{HW} \sum_{=1} \sum_{=1} A^{(\,)}(i, j). \quad (1)$$

Grad-CAM extends CAM to architectures not limited by global average pooling. [23, 43] Let $S$ denote the logit for class $c$, and $A$ the $k$-th feature map of a target convolutional layer. The importance weight of feature map $k$ is then derived from the gradients:

$$\alpha = \frac{1}{HW} \sum_{=1} \sum_{=1} \frac{\partial S}{\partial A(i, j)}. \quad (2)$$

The unnormalized class activation map is then generated by applying a weighted sum over the feature maps, followed by a ReLU activation:

$$L_{\text{Grad-CAM}}(i, j) = \text{ReLU}\left( \alpha\, A(i, j) \right). \quad (3)$$

In our pipeline, Grad-CAM is applied to each convolutional block to quantify *layer-wise relevance*. For layer $\ell$, relevance is summarized by the mean heatmap intensity.

$$m_{\ell} = \frac{1}{HW} \sum_{=1} \sum_{=1} L_{\text{Grad-CAM},\ell}(i, j), \quad (4)$$

which yields a scalar importance score per layer.

## 3.5 XAI-guided architecture refinement

Deep tumor classifiers often increase depth and parameters to improve accuracy, at the cost of interpretability and deployment efficiency. We use explanation evidence to simplify the CNN:

(1) Train the baseline CNN on the preprocessed training data.

(2) Compute $m_{\ell}$ for all convolutional blocks using Eq. (4).

across layers:

$$\tau = \text{Percentile}\,(m_{\ell}) \quad (5)$$

In practice, layers whose relevance scores fall within the lowest percentile of the distribution of $m_{\ell}$ values are considered low-contribution layers and are removed before retraining. This strategy ensures that pruning targets layers that consistently provide minimal explanatory relevance across samples.

(4) Remove layers (blocks) whose importance falls below $\tau$, rebuild the network, and retrain using the same protocol.

This yields a lighter model while promoting attention to tumor-relevant regions.

**Table 2: Summary of recent MRI-based brain tumor classification methods (with and without XAI).**

| Approach | MRIs | Acc. | Explainability | Key caveat |
|---|---|---|---|---|
| ConvAttenMixer [3] | 7023 | 97% | None | High architectural complexity |
| Shallow CNN (7 layers) [14] | 253 | 94% | None | Small dataset; binary setup |
| AlexNet + KNN [2] | 2880 | 97% | None | Moderate data scale |
| Enhancement + custom CNN [34] | 7023 | 97% | None | Heavy model design |
| Revised CNN [30] | 7023 | 96% | None | Below top-performing baselines |
| Pretrained InceptionV3 [13] | 7023 | 97% | None | Capacity/compute overhead |
| Deeper CNN (24 layers) [32] | 3264 | 94% | None | Lower accuracy |
| VGG-based CNN [15] | 4200 | 96% | None | Complex training pipeline |
| MobileNet (transfer learning) [16] | 7023 | 99% | None | Relatively complex backbone |
| CNN features + KNN [36] | 2879 | 95% | None | Accuracy gap to best models |
| Fuzzy twin SVM [4] | 150 | 93% | None | Tuning-sensitive; weaker results |
| Deep CNN baseline [6] | 30 | – | None | Metrics not consistently reported |
| ViT ensemble [38] | 3064 | 98% | None | High compute demand |
| Fine-tuned DenseNet [7] | 3064 | 97% | None | Limited dataset scope |
| VGG19 + Grad-CAM [20] | 253 | 98% | Grad-CAM | Small dataset; binary setup |
| VGG16 + SHAP [8] | 3000 | N/A | SHAP | Binary; accuracy not stated |
| VGG16 + LRP [1] | 3000 | 97% | LRP | Binary setup |
| Dual-input CNN (LIME/SHAP) [10] | 2870 | 85% | LIME, SHAP | Lower generalization |
| ResNet50 + Grad-CAM [27] | 3000 | 98% | Grad-CAM | Binary setup; limited scale |

## 3.6 Explainability validation with SHAP and LIME

We employ complementary explainers to corroborate Grad-CAM evidence. SHAP attributes feature contributions using Shapley values from cooperative game theory [11, 25]. For local, instance-level verification, we apply LIME, which perturbs an input and fits an interpretable surrogate model around the neighborhood of the instance [21, 33]. LIME weights perturbed samples $z$ by proximity to the original input $x$:

$$\pi(z) = \exp\left(-\frac{D(x, z)^2}{\sigma^2}\right), \tag{6}$$

where $D(\cdot, \cdot)$ denotes a distance measure and $\sigma$ controls locality. In our study, Grad-CAM is the primary method due to its spatial localization suitability for MRI, while SHAP and LIME provide additional validation from attribution and local-surrogate perspectives.

# 4 Results and Discussion

## 4.1 Experimental Protocol

The experiments were carried out in Python 3 with TensorFlow on Kaggle Notebooks, using an NVIDIA Tesla P100 GPU (16 GB) and 13 GB of RAM. For Dataset-1, we used an 80%/10%/10% split, resulting in 5,618 training images, 702 validation images, and 702 test images. Dataset-2 was kept separate and used only to assess out-of-domain generalization.

Model settings were determined through empirical exploration, with guidance from earlier studies. We examined learning rates of {0.01, 0.005, 0.001, 0.0005}, and selected 0.001 because it provided the most consistent convergence behavior. We also tested batch sizes of {16, 32, 40, 64}, where 40 achieved a favorable balance between computational efficiency and training stability. The models were trained for 40 epochs, since extending training beyond this point produced no noticeable improvement.

From a computational perspective, the baseline CNN completed training in 86 s over 24 epochs and required, on average, 10 ms to process a single image during inference. The proposed XAI-enhanced pipeline took 135 s for 40 epochs, with inference increasing slightly to 11 ms per image. To analyze model interpretability, we employed Grad-CAM, SHAP, and LIME.

Classification performance was evaluated using Accuracy, Precision, Recall, and F1-score, based on true positives (TP), true negatives (TN), false positives (FP), and false negatives (FN):

$$\text{Accuracy} = \frac{\text{TP} + \text{TN}}{\text{TP} + \text{TN} + \text{FP} + \text{FN}}, \tag{7}$$

$$\text{Precision} = \frac{\text{TP}}{\text{TP} + \text{FP}}, \tag{8}$$

$$\text{Recall} = \frac{\text{TP}}{\text{TP} + \text{FN}}, \tag{9}$$

$$\text{F1} = \frac{2 \times \text{Precision} \times \text{Recall}}{\text{Precision} + \text{Recall}}. \tag{10}$$

## 4.2 Explainability analysis

We first compare the Grad-CAM visualizations of the baseline model and the proposed XAI-guided model, and then further examine the highlighted evidence using SHAP and LIME. Overall, the proposed model shows more focused and clinically meaningful attention, indicating less dependence on irrelevant or non-discriminative regions.

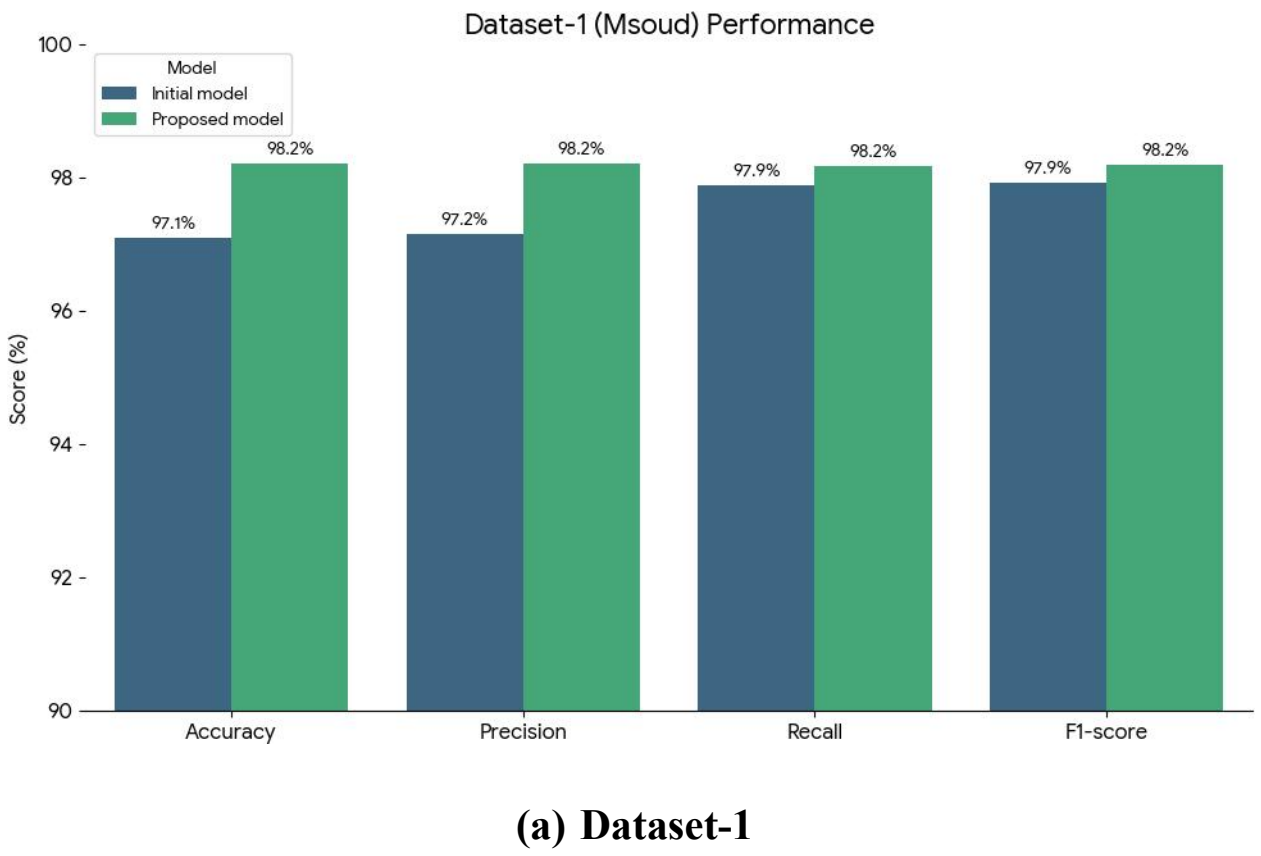

**(a) Dataset-1**

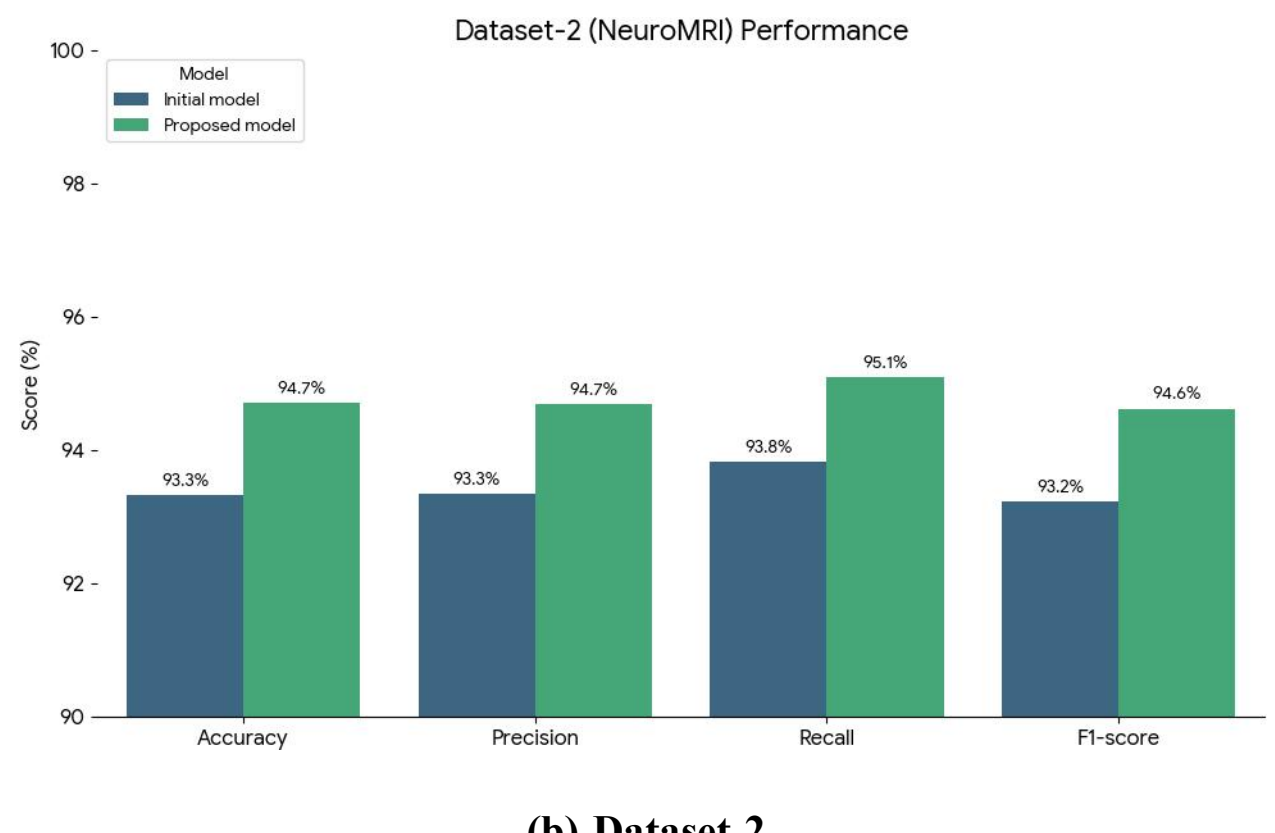

**(b) Dataset-2**

**Figure 2: Bar charts illustrating classification performance on Dataset-1 and Dataset-2.**

**Table 3: Overview of the hyperparameters used for model training.**

| Parameter | Value |
|---|---|
| Optimizer | Adam |
| Learning rate | 0.001 |
| Loss function | Sparse categorical cross-entropy |
| Batch size | 40 |
| Epochs | 40 |

**Grad-CAM.**Grad-CAM highlights the image regions that contribute most strongly to the predicted class. In meningioma cases, the baseline model often shows diffuse attention, whereas the proposed model more consistently focuses on extra-axial regions adjacent to the meninges, which is more consistent with the typical anatomical origin of these tumors. For glioma samples, the baseline model sometimes extends attention into surrounding tissue, while the proposed model concentrates more clearly on intra-axial tumor regions, including irregular boundaries and central low-intensity areas that may correspond to necrotic components in higher-grade lesions. In pituitary tumor cases, the baseline model occasionally attends to nearby anatomical structures, whereas the proposed model primarily highlights abnormal tissue in the sellar region, in line with clinical expectations. These qualitative observations suggest that the XAI-guided refinement strategy promotes greater reliance on tumor-relevant evidence.

**SHAP and LIME.** SHAP heatmaps provide attribution cues by highlighting regions that contribute positively (high importance) or negatively (low importance) to a prediction. Across classes, SHAP attributions align with disease-specific morphology: glioma predictions are supported by irregular borders and central heterogeneous regions; meningioma predictions emphasize extra-axial areas along the meninges; and pituitary predictions highlight the sellar region and nearby anatomical context. LIME further supports these findings with instance-level superpixel explanations. It highlights infiltrative regions in gliomas, dural attachment patterns in meningiomas, and the sellar region in pituitary tumors, suggesting that the proposed model relies on clinically relevant areas rather than background artifacts.

## 4.3 Classification performance and generalization

Table 4 and Fig. 2 present the quantitative comparison between the baseline and the proposed models. On Dataset-1, the proposed approach attains an accuracy of 98.21%, accompanied by consistently strong precision, recall, and F1-score, reflecting effective discrimination across tumor subtypes.To assess generalization, we evaluate the trained model on Dataset-2 without adaptation. The proposed approach attains 94.72% accuracy on this unseen dataset, demonstrating robust performance under domain shift. Relative to the baseline, the proposed model improves all reported metrics on both Dataset-1 and Dataset-2, supporting the effectiveness of the XAI-guided refinement strategy for enhancing classification performance while maintaining generalization.

# 5 Discussion and Comparison

This study introduces an explainability-guided CNN for multi-class brain tumor classification using MRI scans. The baseline network accepts input images of size 224×224×3 and follows a conventional convolutional design composed of repeated *Conv(3×3) + ReLU + MaxPool* blocks, with the number of filters progressively increasing from 8 to 256. These layers are followed by batch normalization, average pooling, and fully connected layers, with softmax used to generate probabilities over four tumor classes. While the baseline model delivers strong classification performance, its prediction mechanism remains insufficiently transparent, making it difficult to determine whether decisions are driven by tumor-relevant patterns or by less informative regions such as surrounding normal tissue.

## 5.1 Explainability-Oriented Refinement and Interpretation

To address this limitation, we integrate Grad-CAM into the training loop and use layer-wise relevance estimates to refine the architecture. Specifically, Grad-CAM is computed across convolutional blocks to quantify each layer's contribution, and layers below a relevance threshold are removed before retraining. This explanation-guided pruning yields a simpler and more interpretable model without sacrificing performance, as the retained layers emphasize features that are most informative for subtype discrimination.

**Table 4: Performance comparison of the baseline and proposed models on Dataset-1 and Dataset-2.**

| Dataset | Model | Accuracy | Precision | Recall | F1-score |
|---|---|---|---|---|---|
| Dataset-1 | Initial model | 97.10% | 97.16% | 97.90% | 97.93% |
| | Proposed model | 98.21% | 98.22% | 98.18% | 98.20% |
| Dataset-2 | Initial model | 92.33% | 93.35% | 93.83% | 93.24% |
| | Proposed model | 94.72% | 94.70% | 95.10% | 94.63% |

We validate interpretability using three complementary explainers: Grad-CAM, SHAP, and LIME. Grad-CAM provides spatial localization of class evidence and shows that, relative to the baseline, the refined model produces more concentrated attention over abnormal tumor regions. SHAP offers attribution-based evidence indicating that the strongest positive contributions are associated with tumor-relevant structures, while LIME provides instance-level verification through localized superpixel explanations. The consistency across these methods strengthens confidence that the proposed model bases predictions on clinically meaningful tissue rather than spurious cues.

## 5.2 Performance and generalization

Quantitatively, the proposed model outperforms the baseline, achieving 98.21% accuracy and F1-score on Dataset-1, demonstrating reliable subtype separation. To assess robustness under domain shift, we evaluate on Dataset-2 (unseen during training) and obtain 94.72% accuracy and F1-score. While some degradation is expected due to differences in acquisition and data distribution, the limited drop indicates strong generalization and supports suitability for real-world use where heterogeneous data are common. Importantly, the observed performance gains are aligned with the explanation results, suggesting that improved accuracy is accompanied by improved evidence quality.

## 5.3 Comparison with prior work

Many recent brain tumor classification methods improve performance by increasing architectural complexity, such as through attention modules, transformer-based designs, or large transfer-learning backbones. In contrast, our findings show that a relatively lightweight CNN can still achieve strong results when interpretability signals are used to guide model refinement. By using Grad-CAM-based layer-wise relevance scores to remove low-contributing layers, the proposed model maintains high classification performance while avoiding unnecessary architectural complexity. This suggests that highly parameterized models may not always be essential for MRI-based brain tumor classification when refinement is driven by clinically meaningful explanation cues.

## 5.4 On statistical testing

We emphasize clinically relevant behavior—whether the model consistently attends to tumor regions—in addition to aggregate metrics. While statistical tests can quantify differences in accuracy, they do not directly capture whether predictions are supported by anatomically plausible evidence. A more extensive statistical analysis across multiple random seeds and additional cohorts remains an important direction for future work.

# 6 Threats to Validity

Our evaluation is limited to MRI; results may not transfer directly to other modalities (e.g., CT or PET) without adaptation. In addition, although we use Grad-CAM, SHAP, and LIME to validate interpretability, other explanation methods may reveal different aspects of model behavior. Finally, the pruning threshold is derived from Grad-CAM relevance statistics and may vary across datasets and training runs; future work should study stability across random seeds and alternative pruning criteria.

# 7 Conclusion and Future Work

We presented an explainability-driven CNN framework for brain tumor classification that uses XAI not only for post hoc interpretation but also for model refinement. By combining standardized preprocessing with Grad-CAM-guided architectural simplification and validating decisions using SHAP and LIME, the proposed model produces more transparent and reliable predictions grounded in abnormal tumor tissue. The proposed approach achieves 98.21% accuracy on Dataset-1 and 94.72% accuracy on the unseen Dataset-2.

Future work will focus on narrowing the gap between experimental performance and real-world clinical deployment. In particular, improving inference speed and memory efficiency will be important for integrating the proposed model into time-sensitive clinical workflows, such as radiology decision-support systems. Another key direction is the quantitative evaluation of interpretability, including how explanation quality affects clinician trust, diagnostic confidence, and willingness to rely on model outputs in practice. User-centered studies involving radiologists may further clarify whether explanation-guided visualizations improve transparency and reduce diagnostic uncertainty. Future research should systematically assess robustness across multi-institutional datasets, scanner configurations, and imaging protocols to ensure stable performance under realistic clinical variability. Overcoming these challenges is crucial for advancing explanation-driven deep learning models from experimental research toward dependable clinical deployment.